\documentclass[preprint,12pt]{elsarticle}

\usepackage{amssymb}
\usepackage{fancyvrb}
\usepackage[T1]{fontenc}
\PassOptionsToPackage{hyphens}{url}
\usepackage[hidelinks]{hyperref}
\usepackage{todonotes}
\usepackage{siunitx}
\usepackage{lineno}
\modulolinenumbers[5]
\usepackage{xcolor}
\usepackage{amsmath}
\usepackage{cleveref}
\usepackage{listings}
\usepackage[ruled,vlined]{algorithm2e}

\definecolor{dkgreen}{rgb}{0,0.6,0}
\definecolor{grey}{rgb}{0.5,0.5,0.5}
\definecolor{mauve}{rgb}{0.58,0,0.82}
\newcommand{\pyinline}[1]{\lstinline[postbreak={}]{#1}}
\newcommand{\cpinline}[1]{\lstinline[postbreak={}]{#1}}

\newcommand{\fluka}[1]{\texttt{\MakeUppercase{#1}}}
\newcommand{\PYGEOMETRY}{\textsc{Pyg4ometry}}
\newcommand{\PYGEOMETRYPOS}{\textsc{Pyg4ometry's}} % possessive

\newcounter{bla}

\journal{Computer Physics Communications}

\begin{document}

\begin{frontmatter}



\title{\PYGEOMETRY{}: a Python library for the creation of Monte Carlo radiation transport physical geometries}


\author[a]{S.~D. Walker}
\author[a]{A. Abramov}
\author[a]{L.~J. Nevay}
\author[a]{W. Shields}
\author[a]{S.~T. Boogert\corref{author}}

\cortext[author] {Corresponding author.\\\textit{E-mail address:} stewart.boogert@rhul.ac.uk}
\address[a]{John Adams Institute at Royal Holloway, Department of Physics, Royal Holloway, Egham, TW20 0EX, Surrey, UK}

\begin{abstract}
Creating and maintaining computer-readable geometries for use in Monte Carlo Radiation Transport (MCRT) simulations is an
error-prone and time-consuming task. Simulating a system often requires geometry from different sources and modelling
environments, including a range of MCRT codes and computer-aided design (CAD) tools. \PYGEOMETRY{} is a Python library
that enables users to rapidly create, manipulate, display, debug, read, and write Geometry Description Markup Language (GDML)-based
geometry used in MCRT simulations. \PYGEOMETRY{} provides importation of CAD files to GDML tessellated solids, conversion of GDML geometry
to FLUKA and conversely from FLUKA to GDML. The implementation of \PYGEOMETRY{} is explained in detail in this paper and includes a number of small
examples to demonstrate some of its capabilities. The paper concludes with a complete example using most of \PYGEOMETRYPOS{} features and a discussion of possible
extensions and future work.
\end{abstract}


\end{frontmatter}

% \linenumbers

% Computer program descriptions should contain the following
% PROGRAM SUMMARY.

%\linenumbers

{\noindent \bf PROGRAM SUMMARY}
  %Delete as appropriate.

\begin{small}
\noindent
{\em Program Title: \PYGEOMETRY{} }                                         		\\
{\em Licensing provisions: GPLv3 }							\\
{\em Programming language: Python, C++}                         		\\
{\em External routines/libraries: ANTLR, CGAL, FreeCAD, NumPy, OpenCascade, SymPy, VTK} \\
%{\em Supplementary material:}                                 				\\
  % Fill in if necessary, otherwise leave out.
%{\em Journal reference of previous version:}                  			\\
  %Only required for a New Version summary, otherwise leave out.
%{\em Does the new version supersede the previous version?:}   	\\
  %Only required for a New Version summary, otherwise leave out.
%{\em Reasons for the new version:}							\\
  %Only required for a New Version summary, otherwise leave out.
%{\em Summary of revisions:}*								\\
  %Only required for a New Version summary, otherwise leave out.
{\em Nature of problem:}\\  % (approx. 50-250 words)
Creating computer-readable geometry descriptions for Monte Carlo radiation transport (MCRT) codes is a time-consuming and error-prone task.
Typically these geometries are written by the user directly in the file format used by the MCRT code. There are also multiple MCRT codes
available and geometry conversion is difficult or impossible to convert between these simulation tools.
\\
{\em Solution method:}\\  % (approx. 50-250 words)
Create a Python application programming interface for the description and manipulation of Geant4 and FLUKA geometries, with full support for
the direct reading and writing of their respective geometry description file formats.
Form triangular meshes to represent geometric objects for both visualisation of the
geometry and to enable the use of advanced mesh-based geometric algorithms. Triangular mesh processing algorithms allow the loading and use of STL and CAD/CAM
files. Converting from FLUKA to Geant4 requires algorithms to decompose solids to a set of unions of convex solids. Converting from
FLUKA to Geant4 requires a number of steps including the replacement of infinite surfaces with finite solids and the automatic elimination of overlaps.

%{\em Additional comments including Restrictions and Unusual features (approx. 50-250 words):}\\
  %Provide any additional comments here.

%\begin{thebibliography}{0}
%\bibitem{1}Reference 1         % This list should only contain those items referenced in the
%\bibitem{2}Reference 2         % Program Summary section.
%\bibitem{3}Reference 3         % Type references in text as [1], [2], etc.
                               % This list is different from the bibliography at the end of
                               % the Long Write-Up.
%\end{thebibliography}
%* Items marked with an asterisk are only required for new versions
%of programs previously published in the CPC Program Library.\\
\end{small}

\section{Introduction} \label{sec:introduction}
There are numerous different software codes to simulate the passage of particles through material, such radiation transport (RT) programs
include MCNP~\cite{Mcnp_Werner}, FLUKA~\cite{Fluka_Ferrari,Fluka_Bohlen}, Geant3~\cite{Geant3_Brun} and Geant4~\cite{Geant4_Agostinelli}.
All these codes are based on the Monte Carlo technique but each code either has a particular specialism, simulation methodology or target user community.
 Monte Carlo RT (MCRT) simulations have diverse uses including shielding calculations for radiological protection, detector performance, medical
imaging and therapy, and space radiation environment simulations. A fundamental requirement of all of the codes is to supply a
computer-readable description of the  physical 3D geometry that  the particles are passing through.  The creation of geometry files is
typically a very time-consuming activity and the simulation validity and performance is directly dependent on the quality of the geometry. There is no
standard geometry format used across MCRT codes, with each code often using its own unique format. A user will typically not have geometry in both FLUKA and Geant4 for example. A geometry system
that enables the conversion between files prepared for different codes will allow for cross-checks of the physics processes in different particle transport
 codes.  The file formats used for geometry are generally focused
 on the computational efficiency of particle tracking algorithms  and not ease of preparation. In addition to the creation of geometry files for RT programs, usually
 computer-aided design (CAD) files exist for systems which need to be simulated. The fundamental geometric representations in CAD files are usually not
 amenable to MCRT programs.  For these reasons it is advantageous to create a software tool that allows particle transport code users to rapidly develop
 error-free geometry files, convert between common MCRT geometry formats and load CAD models.

\begin{figure*}[hbt!]
  \normalsize
  \centering
  \includegraphics[width=0.7\textwidth]{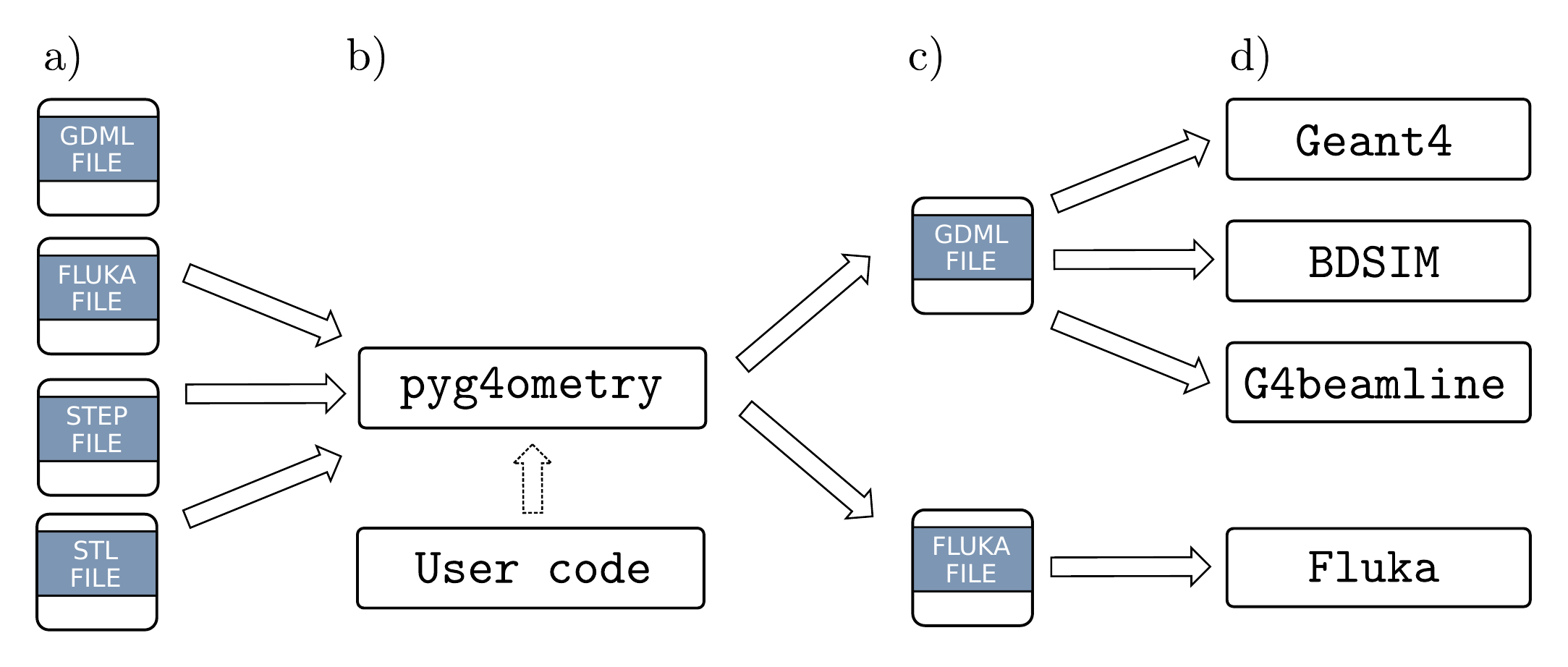}
  \caption{\label{fig:workflow}Schematic of \PYGEOMETRY{} workflow, showing (a) the different input file formats, (b) Python processing, (c) output file targets and
  (d) MCRT codes which use the geometry. }
\end{figure*}

This paper describes a geometry creation and conversion package called \PYGEOMETRY{}, written in Python and internally based on the Geant4 application
programming interface (API) and the Geometry Description Markup Language (GDML) for file persistency~\cite{GDML}. The main features of \PYGEOMETRY{}
are a Python scripting API to rapidly design parametrised geometry; conversion to and from  FLUKA geometry descriptions; conversion from CAD formats  (STEP
and IGES) based on FreeCAD~\cite{FreeCAD} and OpenCascade~\cite{OpenCASCADE}; and powerful geometry visualisation tools based on VTK~\cite{VTK4}. The
origin of \PYGEOMETRY{} was a set of utilities to prepare geometry for an accelerator beamline simulation program based on Geant4 called BDSIM~\cite{BDSIM_Nevay}.
Accelerator physicists, like specialists in other areas, need a tool to quickly model specialist geometry and the subsequent interaction of the charged particle beam.
\PYGEOMETRY{} allows the rapid creation and adaptation of geometry, with Figure~\ref{fig:workflow} demonstrating various possible workflows. \PYGEOMETRY{}
is not an executable software package but a toolkit, a user would typically write a very small Python program to use the classes and functions provided by
\PYGEOMETRY{}. This paper describes version 1.0 of \PYGEOMETRY{}, which is freely available as a Git repository and via the Python Package Index (PyPi).

There are existing codes that have functionality similar to \PYGEOMETRY{}. ROOT~\cite{Brun:1997pa}, the high-energy physics data analysis framework,
can load, display and manipulate GDML-like geometry. Existing open source tools to convert CAD files to GDML include GUIMesh~\cite{GUIMesh_Pinto}, with commercial
solutions including ESABASE2~\cite{ESABASE2} and FASTRAD~\cite{FASTRAD}. There are also tools from the fusion and neutronics community that can convert CAD
geometry into formats usable by MCRT codes, with examples including DAGMC~\cite{DAGMC} and McCAD~\cite{McCad}. In principle CAD software can export shape data to STL
(or other similar mesh formats), which can be used by Geant4~\cite{poole2012acad}. However, in practice using a lot of CAD models is difficult if that model is composed of a
large number of parts. This is because exporting, assigning material to, and placing the STL components into the MCRT code can be very cumbersome. Almost all modern CAD tools such as
CATIA, Inventor and SolidWorks have a scripting language to allow users to programmatically generate geometry. Similar CAD-style scripting languages do not exist for either GDML
or FLUKA and the existent set of software does not provide a complete set of tools to efficiently create complex geometries for these codes.

This paper is structured as follows, first a brief radiation
transport-focused introduction to computer descriptions of geometry and an explanation of the design
and implementation of \PYGEOMETRY{}. Subsequent sections describe how \PYGEOMETRY{} can be used to perform rapid geometric modelling as well as conversions from
FLUKA to GDML, GDML to FLUKA and CAD to GDML. The paper concludes with an example of a composite, complex system consisting of components drawn from
all the supported  geometry input files formats.

\section{Computer descriptions of geometry} \label{sec:geometric}
Central to a computer-readable geometry is how a solid is defined in three dimensions. There are numerous different ways to describe a
geometry, including constructive solid geometry (CSG), boundary representation (BREP) and tessellated polygons, which are described
briefly in this section.  The Geant4 geometry specification is a mixture of all three of these geometry modelling techniques and described in detail last.

%%%%%%%%%%%%%%%%% Constructive solid geometry (CSG)
Constructive solid geometry uses Boolean operations (subtraction, intersection and union) between simple solid shapes (e.g. cube, cylinder, sphere, etc.) or infinite
volumes (e.g. a plane-defined infinite half-space) to model complex surfaces which represent a solid. Boolean operations and solids can be combined to form a
CSG tree to model complex geometry. FLUKA uses CSG to model solids, however the form of Boolean expression used by FLUKA is not a general CSG tree but a
logical expression in disjunctive normal form.

%%%%%%%%%%%%%%%%% Boundary representation (BREP)
Boundary representation consists of two parts, topology and geometry. Topological elements are faces, edges and vertices and the corresponding
geometrical elements are surfaces, curves and points. No current MCRT applications use native file formats employed by CAD systems. The conversion of
CAD BREP formats for loading in MCRT applications is typically performed via a tessellated format, although it is possible to decompose BREP descriptions
 to bounded or infinite mathematical surfaces and subsequently solids as used in CSG descriptions. This type of conversion is complex and error-prone,
 although recent progress has been made~\cite{WangNuclSciTech31-82-2020}.

%%%%%%%%%%%%%%%%% Tessellated polygons
Solid volumes can be defined using triangular, quadrilateral or tetrahedral meshes. Numerous formats exist to describe meshes, the ubiquitous being STL with
more modern examples including PLY and OBJ. For solids with curved faces a tessellated mesh will always give an approximate description. As the mesh
deviation distance from the solid decreases the number of polygons increases and with it the memory consumption and execution time of the MCRT simulation.

%%%%%%%%%%%%%%%%% Geant4 geometery
Geant4 geometry description is the richest and most flexible of MCRT codes
and consists of a mixture of BREP, CSG and tessellated concepts. Geant4
includes 27 basic solids and it does not store a sense of topology
present in traditional CAD BREP systems. One of the fundamental solids is a
tessellated solid which can be used to represent STL or PLY files.  Geant4
also provides the ability to perform Boolean operations on these primitive
solids. Not only do solid objects need to be defined but also placed in a
world coordinate system. Geant4 has two concepts which facilitate this:
\emph{logical volumes} and \emph{physical volumes}. A logical volume is a
region of space that is defined by an outer solid but also other attributes
like material, magnetic field and zero or more \emph{daughter} physical
volumes. A physical volume is a unique placement (or instance) of a logical
volume. This design permits large reuse of objects, minimising memory
footprint for largely repetitive structures such as detectors that Geant4
was created to simulate.

%%%%%%%%%%%%%%%%% Volume bounding
If a user is creating and placing multiple daughter volumes within a \emph{mother} volume then it is the user's responsibility to create a solid which fully
encompasses the daughter volumes. Overlaps between daughter volumes and the mother can be detected, but it is desirable
to have a mother volume shape that efficiently holds its daughters. However, there are exceptions to this rule in the form of assembly volumes.

%%%%%%%%%%%%%%%%% FLUKA geometry

%%%%%%%%%%%%%%%%% Geant Detector Mark-up Language
To exchange geometry descriptions between software packages the Geometry Description Markup Language (GDML) was developed~\cite{GDML}.
GDML is an XML-based description of Geant4 geometry. Geant4 and ROOT~\cite{fons_rademakers_2019_3895860} can read and write
GDML and it is commonly used as an exchange format for Geant4 geometries.

\section{\PYGEOMETRY{} design and layout}
\PYGEOMETRY{} is a Python package consisting of semi-independent sub-packages. The sub-package \verb|pyg4ometry.geant4| contains all classes for
Geant4 detector construction and \verb|pyg4ometry.gdml| provides the functionality for reading and writing GDML files. There are sub-packages for importing and
exporting other geometry formats: \verb|pyg4ometry.fluka|, \verb|pyg4ometry.stl| and \verb|pyg4ometery.freecad|.  Lastly, the sub-package \verb|pyg4ometry.convert|
is used for conversions between formats.

The core of \PYGEOMETRY{} consists of Python classes that mimic Geant4 solids, logical volumes, physical volumes, GDML parameters and material classes.
The constructors of the Python classes are kept as close to the original Geant4 C++ implementation as possible so that \PYGEOMETRY{} users do not have to learn
a new API. For example the \verb|G4Box| class in Geant4, has the XML tag \verb|box| in GDML and is represented by the \pyinline{Box}
class in \PYGEOMETRY{}. The Python object initialisers are very similar to their corresponding Geant4 C++ constructors, but the length definitions are those used by GDML. For example,
GDML uses full-lengths whilst Geant4 uses half-lengths. Geometry construction in Python proceeds in a way which is very similar to geometry construction in Geant4.
A user relatively familiar with Geant4 should be able to start creating geometry in \PYGEOMETRY{} immediately. In the rest of this section novel or important developments
in \PYGEOMETRY{} are described.

For each input format supported by \PYGEOMETRY{}  (GDML, STL, FLUKA and STEP) a dedicated \emph{Reader} class is implemented: \verb|gdml.Reader|,
\verb|stl.Reader|, \verb|fluka.Reader| and \verb|freecad.Reader|. Each reader constructs the appropriate \PYGEOMETRY{} classes and provides a \verb|Registry| instance which
can be used or manipulated by the user.
Output consists of taking the registry and writing to file with the desired format.

%%%%%%%%%%%%%%%%%  Internal data representation
The internal data representation closely follows the structure of GDML. A \verb|Registry| class aggregates Python ordered dictionaries that are  used to store the main
elements of a GDML file. As a \PYGEOMETRY{} user instantiates the geometry the associated registry (typically provided as a keyword argument to the class initialiser) is updated. When a user is finished with the geometry, the registry
can be written to disk as a GDML file. It is also possible to modify \verb|Registry| instances, for example by adding or removing volumes, or by combining with other instances to form an aggregate.

%%%%%%%%%%%%%%%%%  Expression evaluation
In GDML symbolic expressions can be used to parametrise solids and their placements. These expressions are evaluated when the GDML is loaded into Geant4.
In order to fully replicate the functionality of GDML an expression engine was implemented using ANTLR~\cite{10.5555/2501720}. The GDML is loaded using
standard XML modules and parsed using ANTLR to create an abstract syntax tree (AST).  GDML allows for the definition
and assignment of variables. GDML expressions are not much more complicated than binary operators $+, -, \times, /$ and common trigonometric and special
functions. The AST terminates on either expressions which evaluate to numbers or GDML variables. Internally, all \PYGEOMETRY{}
classes use GDML expressions and not floating-point numbers. Storing internal data as expressions allows for deferred evaluation (or re-evaluation) of
solid parameters and  placements. This allows a user to update variables whilst defining geometry and the expression engine will update all internal values.
An example of GDML expressions is shown in Listing~\ref{lst:gdmlExpressions}.

\begin{lstlisting}[caption={A simple Python script using \PYGEOMETRY{} to create GDML variables.},label={lst:gdmlExpressions}, language=Python]
# Import modules
import pyg4ometry

# Create empty data storage structure
reg = pyg4ometry.geant4.Registry()

# Expressions
v1 = pyg4ometry.gdml.Constant("v1","0",reg)
v2 = pyg4ometry.gdml.Constant("v2","sin(v1+pi)",reg)

\end{lstlisting}

%%%%%%%%%%%%%%%%%  Replica, division, parametrised and looped volumes
A powerful feature of Geant4 and hence GDML is the ability to either repeat, divide or parametrise geometry. The class which enables the creation of
multiple replicas of a volume in a Cartesian, cylindrical or spherical grid is known as a Replica Volume. A Division Volume breaks a primitive into segments
in either Cartesian or cylindrical polar coordinates. A parametrised volume allows for the arbitrary multiple placement of solids where the parameters are
allowed to vary for each placement.  Another way in GDML to create parametrised solids or volumes is GDML loops, where sections of GDML can be
 repeated with varying parameters based on the loop index. GDML loop loading and expansion are not supported by \PYGEOMETRY{} but will be implemented in a
future release.

\subsection{Tessellation of solids (meshing)}
Creating a uniform 3D mesh description of all solids (including Booleans) is exceptionally useful for visualisation and other algorithms, such as overlap
detection. For each Geant4 solid instance a triangular tessellated vertex-face mesh is generated and cached. This mesh is then used to determine the extent
of placed instances of geometry (physical volumes) and meshes for CSG-derived solids. CSG mesh calculations are performed using a Binary Space Partitioning
(BSP) tree technique in pure Python~\cite{pycsg} or via the Computational
Geometry Algorithms Library (CGAL) surface meshes~\cite{cgal:bsmf-sm-20b}
in C++. The pure Python CSG backend is further accelerated by compiling it
to C using Cython~\cite{cython}.
For maximum flexibility either backend can be used, but in general the
CGAL implementation is one to two orders of magnitude faster than the
Cythonized CSG implementation and should be preferred, particularly for large geometries.
Triangular meshes based on CSG operations involving curved surfaces often contain large numbers of triangles. Before meshes are visualised or written to file
various polygon mesh algorithms  from CGAL~\cite{cgal:lty-pmp-20b} can be employed to give the meshes more desirable features.

\subsection{Visualisation} \label{sec:visualisation}
When implementing geometry a rapid and robust visualisation system is key to produce error-free and efficient simulation input.
A  \PYGEOMETRY{} geometry hierarchy can be viewed using the popular Visualisation Toolkit (VTK). No separate scene graph is required as the Geant4
volume hierarchy is sufficient to place the meshes associated with each physical volume. A daughter volume is placed within a logical volume with a
rotation~$\mathbf{R}_d$, reflection~$\mathbf{S}_d$ and translation~$\mathbf{T}_d$.

\begin{figure}[htb!]
\begin{center}
\includegraphics[width=5cm]{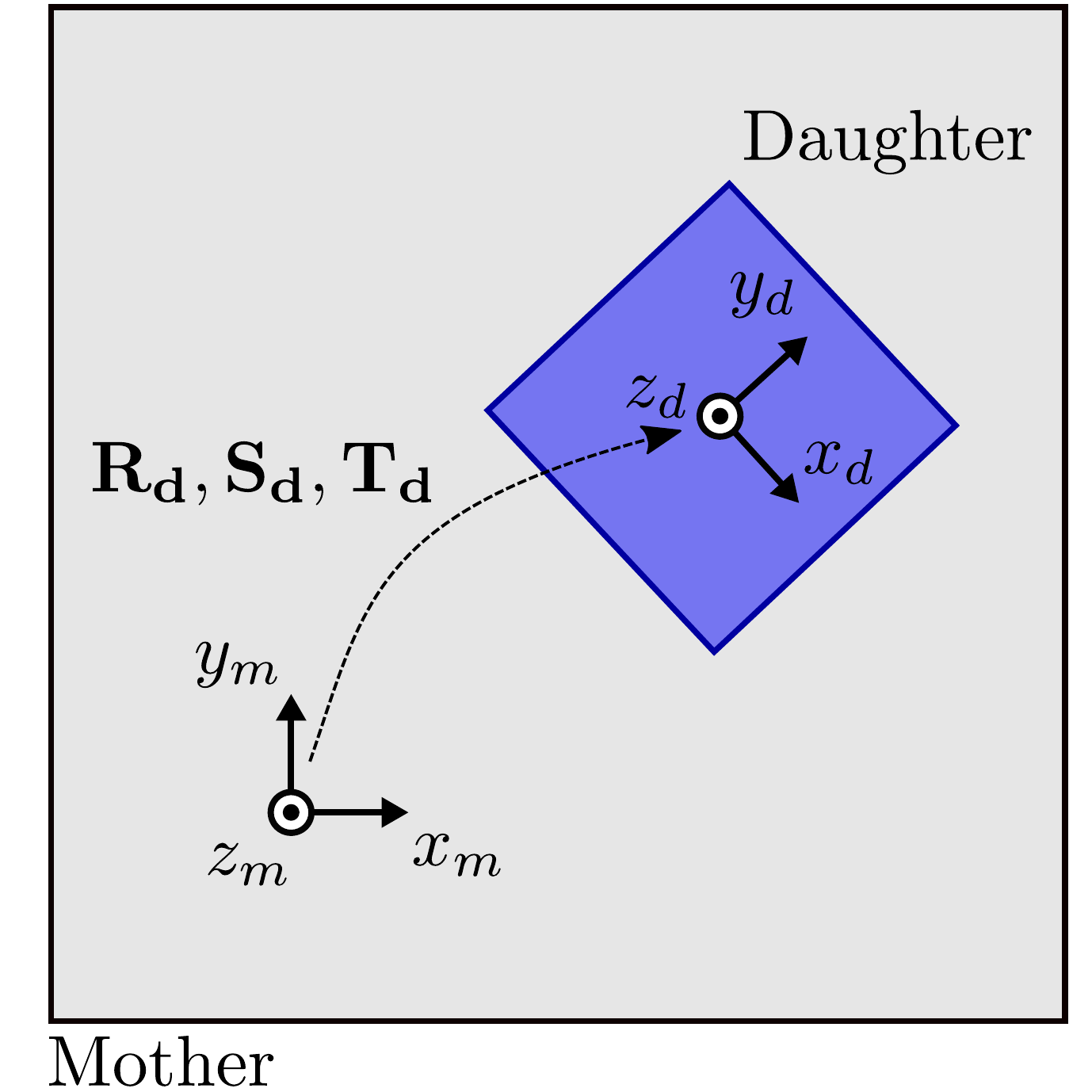}
\caption{The placement of a physical volume inside a logical volume.}
\label{fig:lvToPv}
\end{center}
\end{figure}

The transformation  $\mathbf{M}$  and translation $\mathbf{T}$ from mother to daughter is
\begin{eqnarray}
\mathbf{M} 	& = &  \mathbf{S}_d  \mathbf{R}_d\,, \\
\mathbf{T} 	& = &  \mathbf{T}_d\,.
\end{eqnarray}
If the mother volume is placed in the world then the placement transformation~$\mathbf{M}_w$ and translation~$\mathbf{T}_w$ are expressed as
\begin{eqnarray}
\mathbf{M}_w	  	& = & \mathbf{M}_m \mathbf{M}_d  = \mathbf{S}_m \mathbf{R}_m  \mathbf{S}_d \mathbf{R}_d\, ,				\label{eqn:worldToDaughter1}\\
\mathbf{T}	_w 		& = & \mathbf{M}_m \mathbf{T}_d + \mathbf{T}_m= \mathbf{S}_m \mathbf{R}_m \mathbf{T}_d + \mathbf{T}_m\,,  \label{eqn:worldToDaughter2}
\end{eqnarray}
where the subscript~$m$ indicates mother volume and~$d$ indicates daughter volume. Given a hierarchy of logical and physical volumes,
Equations \ref{eqn:worldToDaughter1} and~\ref{eqn:worldToDaughter2} can be used recursively to place an arbitrary number of nested volumes.

The physical volume class (\pyinline{geant4.PhysicalVolume}) is also used to store visualisation attributes like the solid's
colour, surface or wire-frame representation and visibility. Overlaps detected in the mesh geometry are
stored in the \pyinline{LogicalVolume} instance and can be  displayed separately to allow a user to visually identify and debug the overlaps.

%%%%%%%%%%%%%%%%% Adding properties to volumes
Geometry needs to be augmented with other information for a complete MCRT simulation. Often, other attributes need to be
assigned to regions of space, for example material definition, magnetic field or optical properties. These physical properties
can be used to define the visualisation attributes of a volume.

\subsection{Overlap detection}
\begin{figure}[htbp]
\begin{center}
\includegraphics[width=8cm]{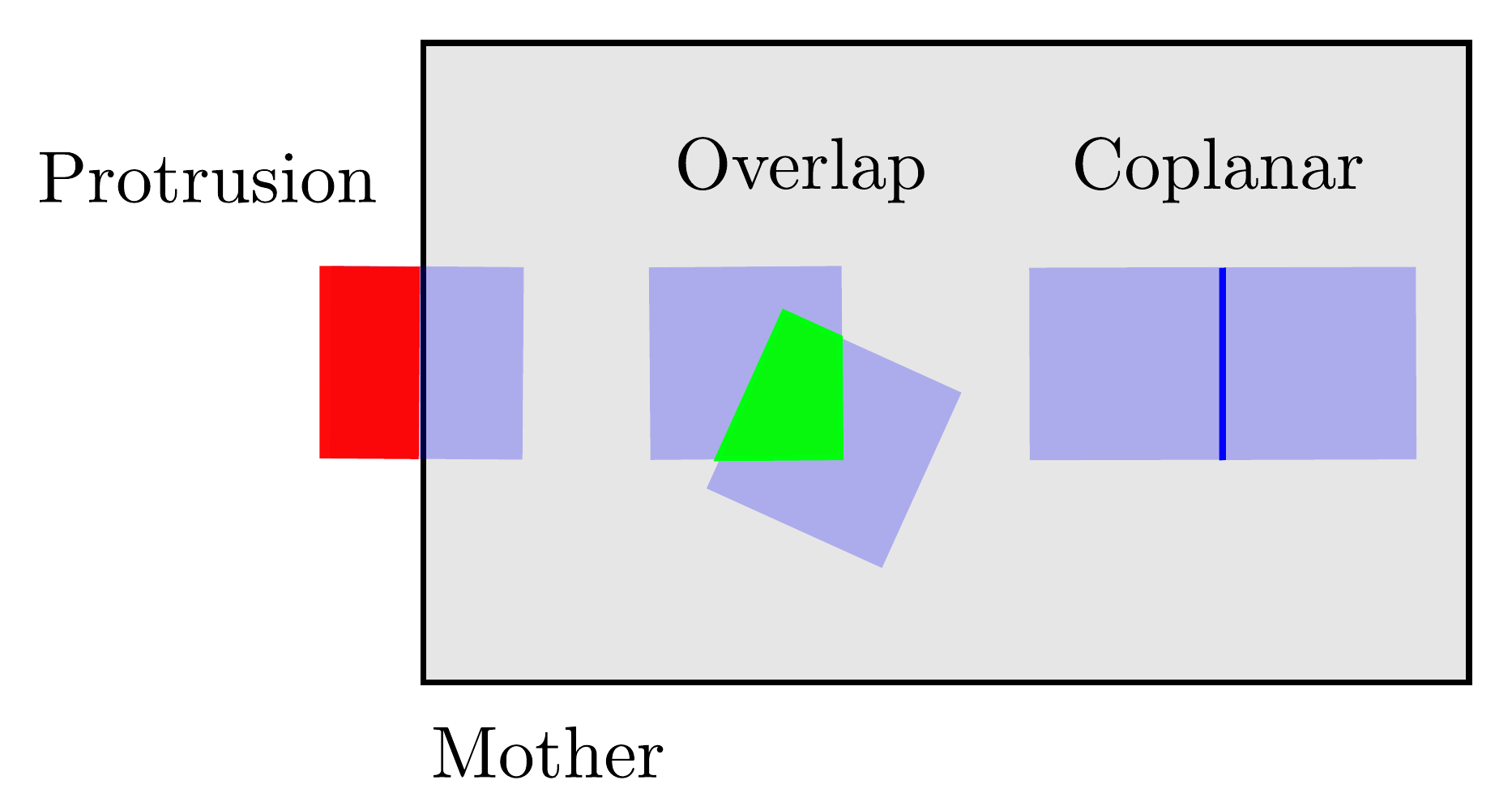}
\caption{Schematic of the three different types of overlap between the daughters of a mother logical volume.}
\label{fig:overlap}
\end{center}
\end{figure}

All MCRT codes cannot handle spatial overlap between two geometric objects and will have ill-defined behaviour when tracking particles
in such a situation.  A key feature of \PYGEOMETRY{} is the detection of potential overlaps in a way which is most useful to the user, it does this by
performing an intersection operation between solid instances and determining if the resulting mesh is empty. Figure~\ref{fig:overlap} shows three different types of possible
overlaps, (a) protrusion of a daughter from the mother, (b)  finite volume intersection between two daughters  and (c) an overlap where two daughters
share a face. If the resulting intersection is non-null then the overlaps can be displayed side-by-side in the visualisation.

\begin{algorithm}[h]
  \SetKwProg{Int}{Function}{}{}
  \SetKwFunction{Intersection}{Intersection}%
  \SetKwProg{OCheck}{Function}{}{}
  \KwData{Logical volume $v$ with mesh $m$ and daughter volume meshes $d\in D$.}
  \KwResult{Set $S$ of non-null mesh intersections.}

  % Define the function used in the algorithm
  \Int{\Intersection{$n_1, n_2$}}{
    \KwData{CSG meshes $n_1$ and $n_2$.}
    \KwResult{The mesh intersection of $n_1$ and $n_2$.}
  }
  % The actual algorithm
  $V \longleftarrow \emptyset$\tcp*[r]{Cache tried mesh pairs.}
  $S \longleftarrow \emptyset$\;

  \For{$d_1 \in D$}{
    $p \longleftarrow$ \Intersection{$m,d_1$}\;
    \If{$p$ is not null}{
      $S \longleftarrow S \cup \{p\}$\;
    }
    \For{$d_2 \in D$}{
      \If{$d_1 = d_2$ {\bf or} $(d_2, d_1) \in V$}{
        \textbf{continue}\;
      }
      $q \longleftarrow \Intersection{a,b}$\;
      \If{$q$ is not null}{
        $S \longleftarrow S \cup \{q\}$\;
      }
      $V \longleftarrow V \cup \{(d_1, d_2)\}$\;
    }
  }
  \label{algo:overlap}
  \caption{The overlap checking algorithm employed in \PYGEOMETRY{}.  The
    algorithm proceeds by performing the CSG intersection of pairs of
    daughter volume meshes for a given logical volume.  Non-null intersections
    between daughter volume meshes are treated as overlaps.}
\end{algorithm}

Overlap detection in \PYGEOMETRY{} relies on the meshes generated for each solid.  As
these mesh representations will generally approximate their respective solid,
so to will any overlap detection algorithm be an approximation. Generally, the overlap detection
algorithm proceeds as shown in Algorithm~\ref{algo:overlap}.
For a logical volume with $n_{\rm daughters}$ physical volumes, assuming meshes for solids have an average number of faces $n$, clearly this algorithm
has complexity $\sim \mathcal{O}(n^2 n_{\rm daughters}^2)$. This complexity
can result in a steep computational cost for highly-flat geometry
hierarchies, but overall remains worthwhile considering the potential
waste in large amounts of cluster CPU time if small overlaps are present in the final MCRT simulation. This algorithm clearly favours geometry
descriptions which have a high degree of logical volume reuse, however this is also true of Geant4 as a whole, so the user will likely be inclined to design along such lines regardless.
Due to the discrete nature of triangular meshes it is not possible to have perfect detection
of overlaps, especially when curved surfaces are considered, and in some rare
cases either overlaps maybe missed, or spurious overlaps maybe reported.  However, the density of the meshes
created for the solids is controllabe by the user, meaning that the user
can opt for more precise overlap checking at the cost of greater
computation time.  In general, the overlap detection algorithm can
present the potential overlaps quickly and easily to the user, thus
significantly aiding the design process.

\section{Rapid geometry modelling}
Given the Python scripting interface, expression  and tessellation engines it is possible for a user to rapidly specify the geometrical layout of the RT problem, vary
the parameters of the geometry and visualise it.  When a user has achieved the desired geometry without geometry overlaps, a GDML file can be written
from the internal memory representation. An example of some of the geometry scripting capabilities of \PYGEOMETRY{} is shown in Listing~\ref{lst:pythonRapidModelling}. The structure should be familiar to regular
users of Geant4 or GDML, apart from the new class described in the previous section called the \verb|Registry|. First the \verb|Registry| is created to store all
the \PYGEOMETRY{} objects; followed by constants;  then materials, solids, logical volumes, physical volumes and other properties such as visualisation attributes; and finally the whole geometry can be saved
as a GDML file or visualised using VTK.

\begin{lstlisting}[caption={A simple Python script using \PYGEOMETRY{} to create a simple Geant4 geometry.},label={lst:pythonRapidModelling}, language=Python]
# import modules
import pyg4ometry.gdml as gd
import pyg4ometry.geant4 as g4
import pyg4ometry.visualisation as vi

# create empty data storage structure
reg = g4.Registry()

# expressions
wx = gd.Constant("wx","100",reg)
wy = gd.Constant("wy","100",reg)
wz = gd.Constant("wz","100",reg)
bx = gd.Constant("bx","10",reg)
by = gd.Constant("by","10",reg)
bz = gd.Constant("bz","10",reg)
br = gd.Constant("br","0.25",reg)

# materials
wm = g4.MaterialPredefined("G4_Galactic")
bm = g4.MaterialPredefined("G4_Fe")

# solids
wb = g4.solid.Box("wb",wx,wy,wz,reg)
b  = g4.solid.Box("b",bx,by,bz,reg)

# structure
wl = g4.LogicalVolume(wb, wm, "wl", reg)
bl = g4.LogicalVolume(b, bm, "b", reg)
bp1 = g4.PhysicalVolume([0,0,0],
                        [0,0,0],
                        bl, "b_pv1", wl, reg)
bp2 = g4.PhysicalVolume([0,0,-br],
                        [-2*bx,0,0],
                        bl, "b_pv2", wl, reg)
bp3 = g4.PhysicalVolume([0,0,2*br],
                        [2*bx,0,0],
                        bl, "b_pv3", wl, reg)

reg.setWorld(wl.name) # define world volume

# physical volume vistualisation attributes
bp1.visOptions.color = (1,0,0)
bp1.visOptions.alpha = 1.0
bp2.visOptions.color = (0,1,0)
bp2.visOptions.alpha = 1.0
bp3.visOptions.color = (0,0,1)
bp3.visOptions.alpha = 1.0

# gdml output
w = gd.Writer()
w.addDetector(reg)
w.write("output.gdml")

# visualisation
v = vi.VtkViewer(size=(1024,1024))
v.addLogicalVolume(wl)
v.addAxes()
v.view()
\end{lstlisting}

\begin{figure}[htbp]
\begin{center}
\includegraphics[width=0.9\columnwidth]{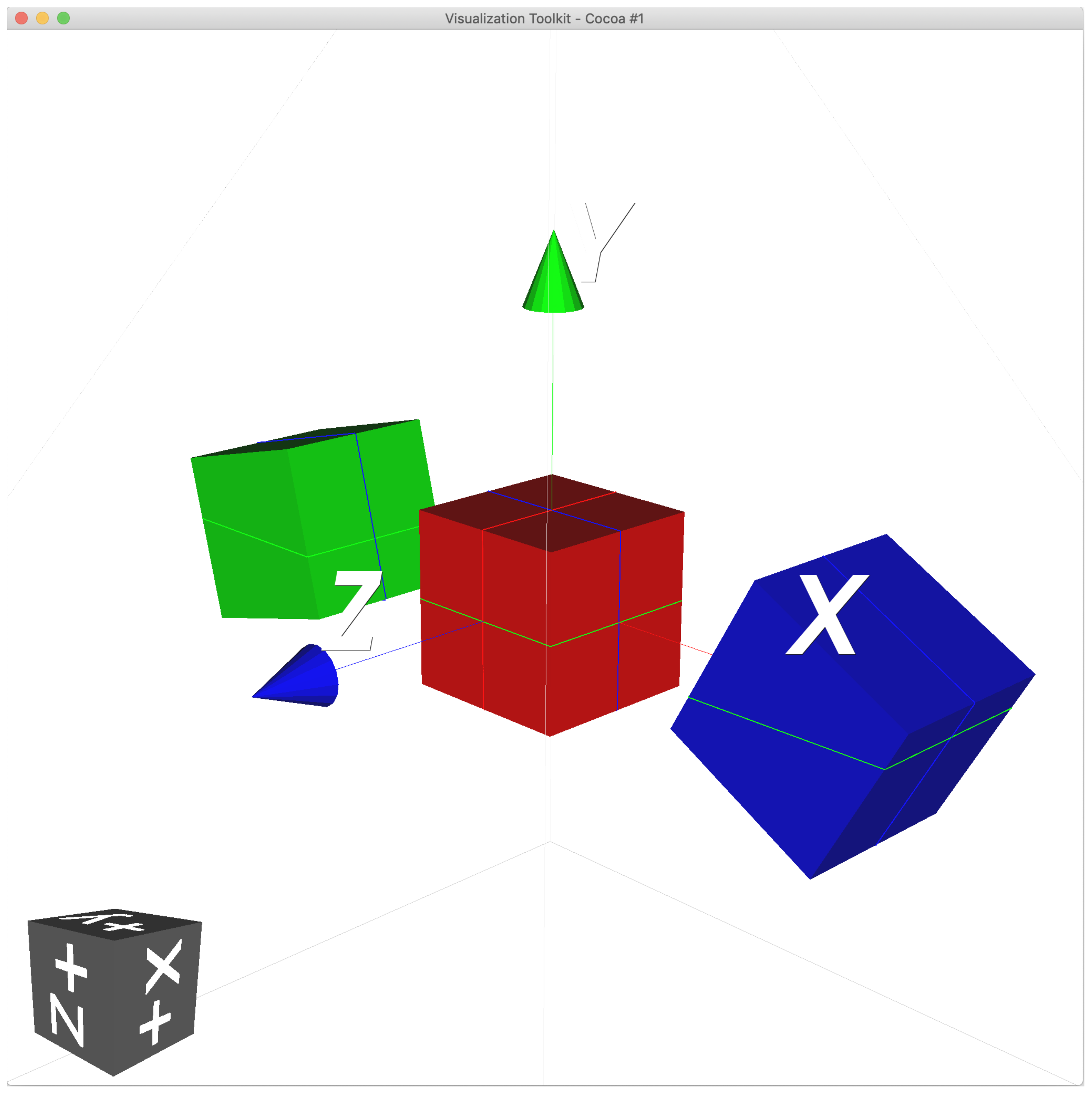}
\caption{VTK visualisation output from code Listing~\ref{lst:pythonRapidModelling}.}
\label{fig:rapidModellingExample}
\end{center}
\end{figure}

An example of the VTK output for code
Listing~\ref{lst:pythonRapidModelling} is shown in
Figure~\ref{fig:rapidModellingExample}. Significantly more complex
geometries can be developed using a structure similar to that shown.

The \PYGEOMETRY{} Python code in the example is approximately as expressive
as the GDML it writes. The benefit of wrapping GDML in Python is that it allows
 very rapid prototyping of geometry without the overhead of C++ compilation  (in the
case of implementing the geometry directly in Geant4) or writing well-formed XML (in the case of GDML). Effectively,
by using \PYGEOMETRY{}, the set of possible user-errors when describing geometry in C++ or XML
instead either manifest as Python exceptions or are eliminated entirely.
Another key benefit is the ability to use the \PYGEOMETRY{} code to create
programmatic converters between different geometry languages or more generally
manipulation and transformation of the geometry stored in memory. The rapid modelling
example given in Listing~\ref{lst:pythonRapidModelling} and Figure~\ref{fig:rapidModellingExample}
is rather trivial, a significantly more complex example is shown in Figure~\ref{fig:gdml-flair}.

\section{FLUKA to GDML conversion}
FLUKA geometry is based upon a limited set of primitives (referred to as
\emph{bodies}) which can be combined using Boolean operations. A
\emph{zone} consists of one or more bodies or \emph{subzones} combined
using intersections and subtractions.  Zones may then be further combined
using union operations to form \emph{regions}, which are defined as the
union of one or more zones, as well as a material.

\begin{figure}[htbp]
\begin{center}
\includegraphics[width=0.9\columnwidth]{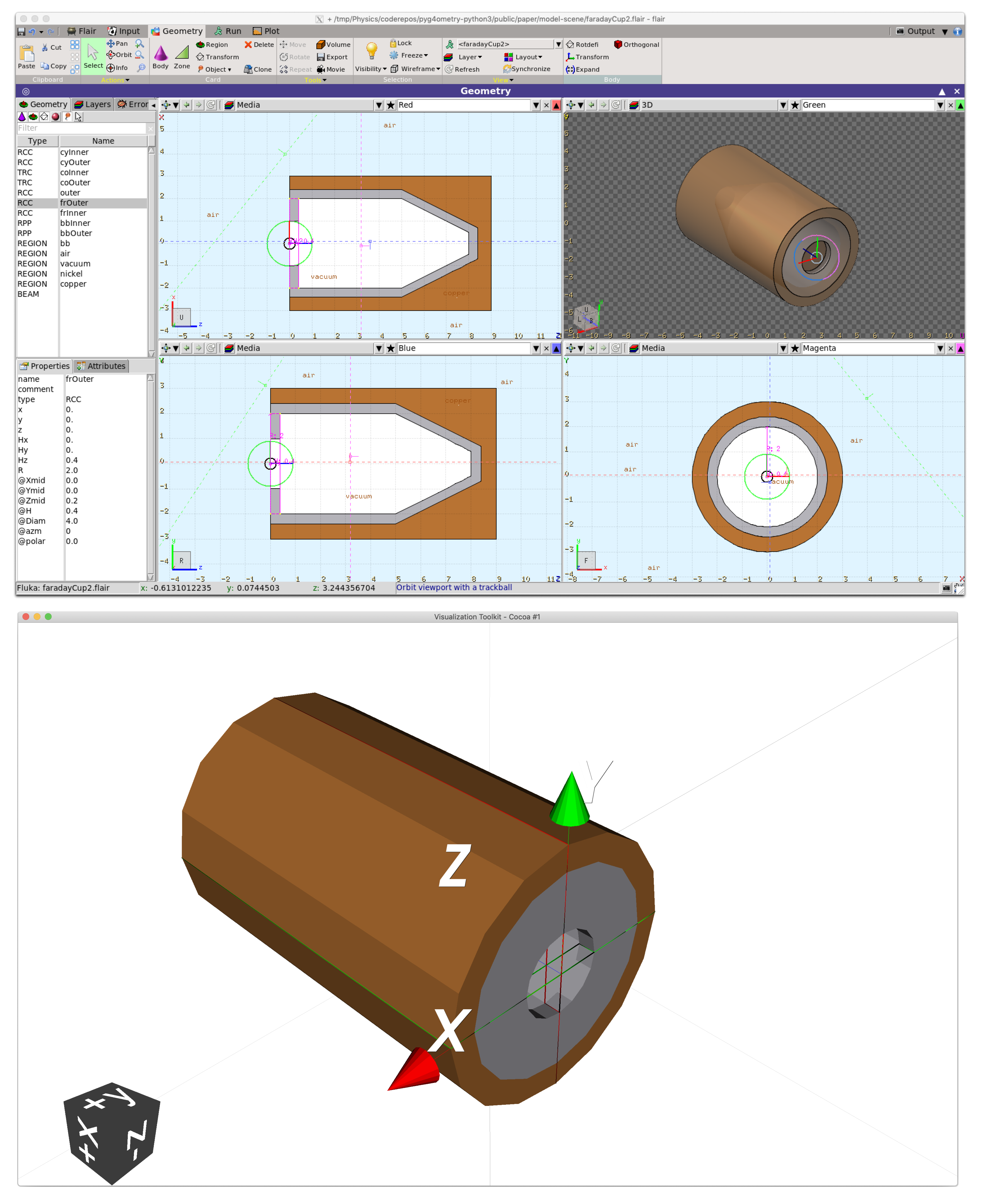}
\caption{Example conversion of a simple FLUKA geometry to GDML. Above:
  the original FLUKA geometry displayed in flair, FLUKA's graphical user interface.
  Below: the GDML geometry viewed using \PYGEOMETRYPOS{} VTK visualiser. The example is a Faraday cup used to capture
and measure accelerator beam charge.}
\label{fig:fluka-to-geant4-cup}
\end{center}
\end{figure}

Each FLUKA body is represented in \PYGEOMETRY{} with a corresponding class,
and in turn each class has a method that returns a GDML primitive solid and
a method that returns that solid's rotation and position such that it
matches its FLUKA equivalent.  The expansion, translation and transform
geometry directives are each folded into one or more of these three
methods.  The mapping of FLUKA bodies to GDML solids is shown in
Table~\ref{tab:Fluka2Geant4}.  It is worth noting that many of the FLUKA
bodies are infinite in extent, but are mapped to finite GDML solids.  The
translation of infinite bodies to equivalent finite solids is one of the
main and most involved steps in the conversion process.  This mapping is
possible because whilst FLUKA bodies can be infinite in extent, all zones
and regions must be finite.  Zones and regions are then composed by
instantiating their respective classes and adding body instances to them.  Each
\pyinline{Zone} and \pyinline{Region} instance can then return its
equivalent GDML Boolean solid.

The FLUKA CSG ASCII is parsed using an ANTLR4-generated parser,
producing an AST.  The resulting AST is then inspected sequentially
(\emph{walked}) to populate \pyinline{Region} instances with zones and bodies.
With the \pyinline{Region} instances populated they can then be manipulated
and translated into GDML.  The translation involves a number of special
steps to bridge the two disparate formats and ensure the resulting GDML is
well-formed and usable in Geant4.  Some of these steps are simple, for
example in FLUKA unions can be disconnected, but in Geant4 specifically
only multi-unions can be disconnected.  Therefore, multi-unions are used
throughout the converted geometry instead of the more conventional binary unions.
Other procedures are more involved and are discussed in the rest of this
section.

\begin{lstlisting}[caption={A simple \PYGEOMETRY{} Python script to load a
    FLUKA file and convert its geometry to a Geant4 logical volume.},label={lst:pythonFlukaLoading}, language=Python]
import pyg4ometry.fluka as fluka
from pyg4ometry.convert import fluka2Geant4

reader     = fluka.Reader("FlukaFileName.inp")
g4Registry   = fluka2Geant4(reader.flukaregistry)
logical      = g4Registry.getWorldVolume()
\end{lstlisting}

\begin{table}[hbt!]
\caption{FLUKA bodies and their corresponding \PYGEOMETRY{} classes.} \label{tab:Fluka2Geant4}
\centering
\begin{tabular}{ll} \hline
FLUKA body                                              & \PYGEOMETRY{} class \\ \hline
RPP (Rectangular parallelepiped)			& Box \\
BOX (General rectangular parallelepiped)		& Box \\
SPH (Sphere)    					& Orb \\
RCC (Right circular cylinder)				& Tubs \\
REC (Right elliptical cylinder)				& EllipticalTube \\
TRC (Truncated Right Angle Cone)			& Cons \\
ELL (Ellipsoid of Revolution) 				& Ellipsoid \\
WED/RAW (Right Angle Wedge)		        	& ExtrudedSolid \\
ARB	(Arbitrary Convex Polyhedron)			& TessellatedSolid \\
XYP 	($X$-$Y$ Infinite half-space)			& Box \\
XZP 	($X$-$Z$ Infinite half-space)			& Box \\
YZP 	($Y$-$Z$ Infinite half-space)			& Box \\
PLA (Generic infinite half-space)			& Box \\
XCC ($X$-axis Infinite Circular Cylinder)		& Tubs \\
YCC ($Y$-axis Infinite Circular Cylinder)		& Tubs \\
ZCC 	($Z$-axis Infinite Circular Cylinder)		& Tubs \\
XEC 	($X$-axis Infinite Elliptical Cylinder)		& EllipticalTube \\
YEC 	($Y$-axis Infinite Elliptical Cylinder)		& EllipticalTube \\
ZEC ($Z$-axis Infinite Elliptical Cylinder)		& EllipticalTube \\
QUA (Quadric surface) 					& TessellatedSolid \\ \hline
\end{tabular}
\end{table}

\subsection{Infinite bodies}
The majority of bodies in FLUKA are infinite in extent, and fall
broadly into four categories, half-spaces, infinitely-long cylinders,
infinitely-long elliptical cylinders and quadric surfaces.
Translating these bodies to Geant4 requires generating the equivalent
finite solid whilst retaining the same final finite Boolean shape.
This is achieved with the use of axis-aligned bounding boxes (AABBs),
in which the FLUKA body is translated to a finite solid with 
dimensions slightly larger than the AABB.  The lengths of infinite
(elliptical) cylinders are reduced to finite equivalents with lengths
slightly greater than the bounding box.  Similarly, half-spaces are
reduced to boxes with one face acting as that of the half-space face,
and quadric surfaces are sampled only over the volume denoted by the
AABB.  Furthermore, the positions of these solids are moved as close
to the bounding box as possible whilst retaining an identical final
Boolean geometry.   Resizing and translating all bodies with respect
to the region bounding boxes ensures that an identical region geometry can
be generated, albeit from much smaller constituent bodies.

Generating these bounding boxes over which the bodies should be translated
to GDML involves first evaluating each region with very large \pyinline{Tub},
\pyinline{EllipticalTube} and \pyinline{Box} instances (by default
\SI{50}{\km} in length), such that they are effectively infinite for most
reasonable use cases.  \PYGEOMETRY{}'s CSG meshing is then used to generate
a mesh for each region from which the axis-aligned bounding box can be
extracted.  Each region is then evaluated a second time with respect to its
 bounding box, with all of its constituent infinite solids being
reduced in size as described above.  The implementation is described in
Algorithm~\ref{algo:redundant-halfspace}. This algorithm works robustly for
half-spaces, infinite circular cylinders and infinite elliptical cylinders
as the number of facets belonging to the mesh for such solids is
independent of the size of the solid, so generating the initial mesh from
arbitrarily large solids works well.  Generating a quadric surface over a very large
volume in space whilst retaining topological information is computationally
very expensive, so to resolve this the user must provide the approximate
axis-aligned bounding box of any region in which a quadric is used to
constrain the size of the mesh only to where it is needed.

\begin{algorithm}

  \KwData{FLUKA regions to be converted to GDML.}
  \KwResult{GDML solids equivalent to the FLUKA regions built from
    minimally-sized primitive solids.}

  \SetKwProg{ToGDML}{Function}{}{}
  \SetKwFunction{ToGDMLSolid}{ToGDMLSolid}

  % The function for translating a fluka body to a finite GDML solid.
  \ToGDML{\ToGDMLSolid{$b, a$}}{
    \KwData{FLUKA body $b$ with axis-aligned bounding box $a$.}
    \KwResult{GDML solid equivalent to $b$ bounded by the volume $a$.}
  }

  \SetKwProg{RAABB}{Function}{}{}
  \SetKwFunction{RegionAABB}{RegionAABB}
  % The function for translating a fluka body to a finite GDML solid.
  \RAABB{\RegionAABB{$r$}}{
    \KwData{FLUKA region $r$.}
    \KwResult{Axis-aligned bounding box (AABB) of the FLUKA region.}
  }

  $ B \longleftarrow \emptyset$\tcp*[r]{Map of regions to AABBs.}
  \For{$r$ in regions to be converted}{
    $B[r] \longleftarrow$ \RegionAABB{$r$}
  }

  \tcp{Map of bodies to minimal bounding boxes.}
  $ E \longleftarrow \emptyset$\;
  \For{$b$ in bodies in regions to be converted}{
    \For{$r$ in regions in which body $b$ is used}{
      $E[b] \longleftarrow E[b] \cup $ \RegionAABB($r$)\;
    }
  }

  $ G \longleftarrow \emptyset$ \;
  \For{body $b$ and AABB $a$ in $E$}{
    \tcp{Map the FLUKA bodies to minimal GDML solids by converting with the
    minimal AABB, $a$.}
    $G \longleftarrow $ \ToGDMLSolid{$b, a$}\;
  }
  \For{$r$ in regions to be converted}{
    build the corresponding GDML for $r$ from the set of minimal GDML
    primitives in $G$.\;
  }

  \label{algo:redundant-halfspace}
  \caption{The infinite-body minimisation algorithm employed in the
    conversion of FLUKA to GDML.}
\end{algorithm}

\subsection{Removing redundant half-spaces}
The above algorithm for replacing infinite FLUKA bodies with finite Geant4
solids works well in most cases, but additional care must be taken for
redundant infinite half-spaces.  A redundant infinite half-space is defined
as one which has no effect on the shape of the final Boolean solid.  Whilst
this may be true of arbitrary bodies, it is most problematic for
half-spaces. If the half-space is far away from the region's AABB after the
infinite body reduction has been performed, then it can result in a
malformed Boolean. Such half-spaces are filtered from their respective
regions during the conversion process by calculating the nearest distance
from the centre of the AABB to the half-space face.  If this distance is
greater than the centre-to-corner distance of the AABB, then that
half-space is deemed to be redundant with respect to that region and is
removed from it during conversion.

\subsection{Coplanar faces}
Coplanar faces present a problem during conversion where the faces of two
union components or that of two regions are perfectly coplanar in FLUKA.
Coplanar faces of this nature are ubiquitous in FLUKA and present no
problems to the operation of the program.  However, in Geant4 these will
generally result in tracking errors.  These must be handled robustly to
ensure the resulting geometry is usable in Geant4.  Coplanar faces are resolved
automatically by slightly decreasing the size of every body that is used in
an intersection, and increasing the size of every body used in a
subtraction.  These rules are inverted for nested subtractions.  This
algorithm ensures that all coplanar faces are removed and replaced with
resolvable, small gaps between faces.  This approach works well for
guaranteeing well-formed geometry that is free from tracking errors in Geant4.

\subsection{Materials}

Any useful translation between geometry description formats must also
account for materials, and accordingly \PYGEOMETRY{} correctly translates
FLUKA materials to GDML.  FLUKA materials can be divided into built-in,
single-element and compound materials.  Built-in materials are simply those
that are predefined by FLUKA, single-element materials are described with a
single \fluka{material} card, and compound materials are described with one
\fluka{material} card followed by one or more \fluka{compound} cards.
These three alternatives are represented in \PYGEOMETRY{} with the
\pyinline{BuiltIn}, \pyinline{Material}, and \pyinline{Compound} classes.
Populating a hierarchy of instances using these classes is made more involved due to
the fact that recursively-defined materials in FLUKA input files need not
be defined in a logical order.  Namely, a given compound may be
defined before the materials that it consists of are themselves defined.
To account for this it is necessary to correctly compute the instantiation
order so that the above classes can be instantiated correctly.  To do this
a directed acyclic graph is populated with the materials and their
constituents, after which a topological sort is performed so that the
compound materials are sequenced after their constituents.  Mapping this
set of nested FLUKA materials to GDML materials is then
straight forwards as the two formats are similarly expressive.

\subsection{Lattice}

\begin{figure}[hbtp]
\begin{center}
\includegraphics[width=6cm]{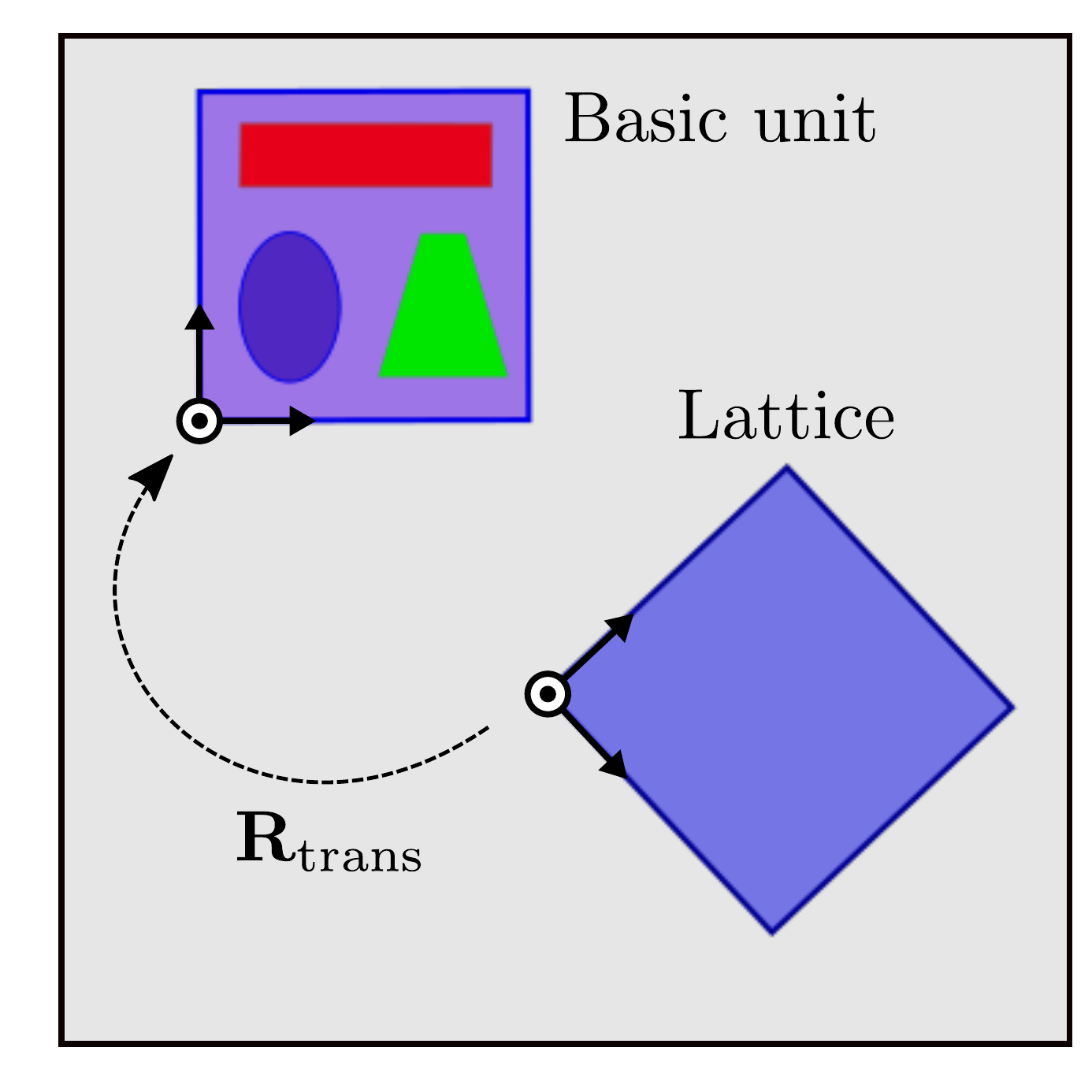}
\caption{The lattice feature demonstrating a lattice cell
  referring to its basic unit with the rototranslation~$R_\textrm{trans}$.
  Any particle entering the cell will be transformed onto the basic unit
  with $R_\textrm{trans}$, and when leaving the basic unit, back to the cell
  with $R_\textrm{trans}^{-1}$.}
\label{fig:lattice}
\end{center}
\end{figure}

FLUKA supports modular geometries with the use of the \fluka{lattice}
command. Figure~\ref{fig:lattice} demonstrates this capability. The
arbitrarily complex \emph{basic unit} can be defined once and used
multiple times by placing one or more empty \emph{lattice cells} with the
associated rototranslation from that lattice cell to the basic
unit. The lattice cells themselves will generally lack structure
and simply serve as a reference to the basic unit. The rototranslated
lattice cell must fully contain the basic unit and all of the regions
within it. When a particle steps into the lattice cell, the
rototranslation is applied to that particle transporting it to the
basic unit, and the simulation continues. Any
particles leaving the basic unit will be transported back to the
lattice cell with the inverted rototranslation. Up to two
levels of lattice nesting are supported in FLUKA.

This feature is clearly analogous to the logical/physical volume feature in
Geant4, although it is more implicit as the contents of a given logical
volume are explicitly stated, whereas the contents of a lattice cell
are implied by the combination of the rototranslation and the global positions of
its lattice cell and basic unit.

Translating the lattice construct into a logical volume (basic unit) with
many physical volumes (lattice cells) requires associating each lattice
cell with the full contents of its corresponding basic unit (typically more
than one region). This is achieved by meshing the complete FLUKA geometry
and then rototranslating the lattice cell mesh with its associated
rototranslation. By construction this will translate the lattice cell mesh
directly onto the full basic unit mesh. This transformed lattice cell mesh
can then be intersected with all other meshes in the scene.  Intersections
with the transformed lattice cell mesh determine the contents of the basic
unit.  Finally, the contents of lattice cell can simply be replaced with a physical
volume containing the logical volumes determined in the previous step.
Thus a basic unit with one or more lattice cells can be translated into a
Geant4 logical volume with one or more physical volumes. As has been
stated, FLUKA supports two levels of lattice nesting, but currently the
conversion to GDML supports only one. However, extending to an extra level
is simple in that it only involves an additional application of a
rototranslation matrix before checking for intersections with the lattice
cell mesh.  This capability will be added in a future release.

\subsection{Discussion}

Figure~\ref{fig:fluka-to-geant4-cup} shows a Faraday cup implemented in
FLUKA and accurately translated to GDML using \PYGEOMETRY{}.  Many of the
steps described above were directly applied in this model and all the
features are tested and demonstrated in the repository.  This set of
algorithms for bridging FLUKA with GDML covers a very broad range of
geometries, however a number of possible improvements remain for the
future.
For example, FLUKA regions can in general be disconnected, that is
to say the resulting set of points defined by that region may consist of
two disconnected subvolumes.  In FLUKA this typically manifests itself in
the form of a disconnected unions, but it is also allowed for intersections
and subtractions to be disconnected.  In the case of translating
disconnected unions from FLUKA to GDML, this presents no difficulties as
the GDML multi-union solid is explicitly allowed to be formed from
disconnected parts.  However, it is forbidden for \pyinline{Intersection} or
\pyinline{Subtraction} solids to be disconnected in this way and is
therefore necessary to account for this difference between the two codes.
One solution to this problem would involve detecting and separating
disconnected meshes, and then placing each separated component as a
separate \pyinline{TesselatedSolid} instance.  This decomposition has not
been implemented and is left as future work.

Furthermore, the quadric surface conversion can be improved by specialising
on some of the individual forms of the quadric.  Simply converting every
quadric into a TesselatedSolid comes at a potential performance cost in the
tracking, as well as usability in \PYGEOMETRY{} as the user must provide an
AABB for every single region featuring a quadric body.  In some cases
tessellation is unavoidable (for example a hyperbolic paraboloid), but
parabolic cylinders, for example, could be mapped to \pyinline{ExtrudedSolid}
instances.
This could provide both a performance improvement in the tracking time and
make \PYGEOMETRY{} easier to use as an AABB would not need to be provided
beforehand.  This has not been implemented as quadrics are relatively
rarely used, but where quadrics are used, it is often in the form of
parabolic cylinders (e.g. magnet pole tips) and this specialisation in
particular would be worth implementing.

\section{GDML to FLUKA conversion}
It is relatively straight forward to convert Geant4 geometry to FLUKA. Each
of the Geant4 solids can be mapped to a FLUKA region.  A region is a volume
of space defined by a material and the Boolean disjunction (a union using
the operator~$\:|\:$ in free format geometry, the most widely-used FLUKA
CSG format) of one or more zones. Each zone is then defined in terms
of the conjunction (intersection with~$+$, subtraction with~$-$) of one or
more primitive bodies, as well as parentheses to determine the order of
operations within the zone.  FLUKA has 20 of these primitive bodies, listed
in Table~\ref{tab:Fluka2Geant4} and, in general, infinite-extent bodies
have tracking accuracy and efficiency benefits over finite ones. Key for
conversion are XY-, XZ-, YZ-Planes (XYP, XZP, YZP), arbitrary plane (PLA),
Z-axis aligned cylinder (ZCC), Z-axis aligned elliptical cylinder (ZEC),
sphere (SPH), truncated right-angle cone (TRC) and general quadric surface
(QUA). Some solids in Geant4 directly map to a single FLUKA body, others
require the construction of a simple FLUKA CSG tree combining these
primitives. Table~\ref{tab:geant2fluka} lists the mapping between Geant4
solids and the bodies used to compose a FLUKA region.
\begin{figure}
\begin{center}
\includegraphics[width=0.9\columnwidth]{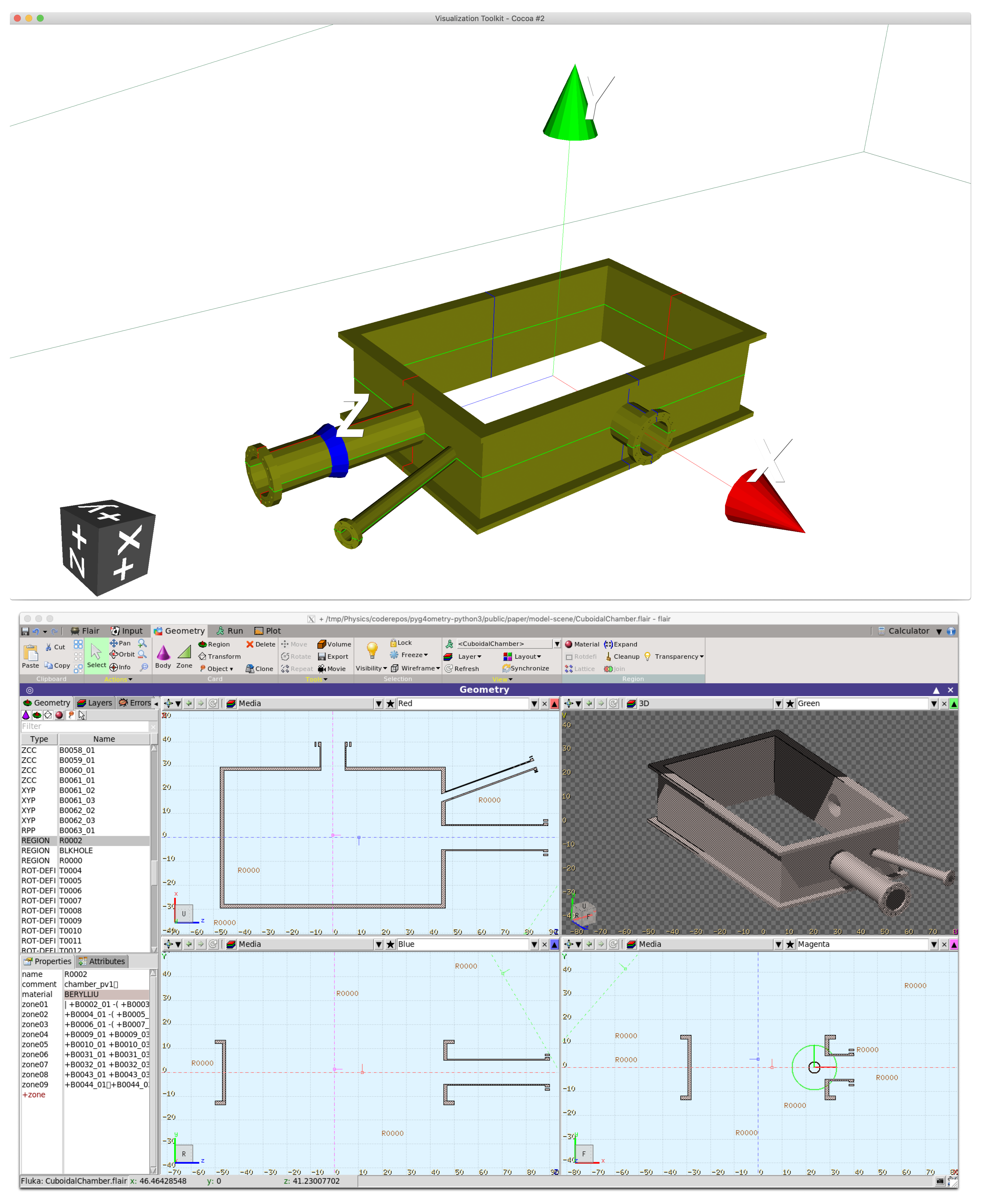}
\caption{Example conversion of a simple GDML geometry to FLUKA, above:
the original model in \PYGEOMETRY{}, below: the converted FLUKA geometry
viewed in flair. The example is a sector bend dipole electromagnet.}
\label{fig:gdml-flair}
\end{center}
\end{figure}

\begin{table}[hbt!]
\centering
\begin{tabular}{ l  l  } \hline
Geant4 solid			& FLUKA region construction		\\ \hline
Box					& +2 XYP + 2 XZP + 2 YZP 		\\
Tube					& +ZCC - 2 PLA -2 XYP - ZCC	 	\\
CutTube				& +ZCC - 4 PLA -ZCC			\\
Cone				& +TRC - TRC - 2 PLA 			\\
Para					& + 6 PLA						\\
Trd					& + 6 PLA						\\
Trap					& + 6 PLA						\\
Sphere				& +SPH - SPH  - 2 PLA - 2 TRC	\\
Orb					& +SPH						\\
Torus				& +$N$ ZCC  - $N$ PLA			\\
Polycone				& +$N$ TRC -2 PLA				\\
Polyhedra				& +$N$ PLA					\\
Eltube				& +ZEC  - 2 XYP				\\
Ellipsoid				& +ELL - 2 XYP		 			\\
Elcone				& +QUA - 2 XYP				\\
Paraboloid			& +QUA - 2 XYP				\\
Hype					& +QUA - QUA - 2 XYP			\\
Tet					& +4 PLA						\\
Xtru					& +$N$ PLA \\
TwistedBox			& +$N$ PLA					\\
TwistedTtap			& +$N$ PLA					\\
TwistedTrd			& +$N$ PLA				 	\\
TwistedTube			& +$N$ PLA					\\
Arb8					& +$N$ PLA					\\
Tessellated			& +$N$ PLA				 	\\
Union				& $R_1 \cup R_2$				\\
Subtraction			& $R_1 - R_2$					\\
Intersection			& $R_1 \cap R_2$				\\
MultiUnion			& $R_1 \cup R_2 \cup R_3 \cup$	\\ \hline
\end{tabular}
\label{tab:geant2fluka}
\caption{GDML/Geant4 solids and the mapping to FLUKA regions.}
\end{table}
Understanding the algebra of FLUKA CSG and how it relates to Geant4's
concepts of solids, logical volumes and physical volumes is necessary
to perform an accurate conversion.  For example, a Geant4 logical volume
solid may correspond to a FLUKA region consisting of many zones
(i.e. unions), e.g.,~$R_1 = +z_1\: |~+z_2$.  If this logical volume is
placed as a physical volume inside some mother volume with its solid
perhaps consisting of multiple zones, e.g., $R_2= +z_3 \: |~+z_4$, then
subtracting a hole in the mother solid to make room for the daughter
results in an equation of the form
\begin{equation}
  \begin{aligned}
R_1 - R_2 	& = & (+z_1 \: | +z_2) - ( +z_3 \: | +z_4) 			\\
			& = & (+z_1 - z_3 - z_4) \;  | \; (+z_2 - z_3 - z_4)\,.
\label{eqn:setDiff}
  \end{aligned}
\end{equation}
Geant4 has Boolean solids associated with difference, union and intersection, so in addition to
Equation~\ref{eqn:setDiff}, both $R_1 \cup R_2$ and $R_1 \cap R_2$ are required in FLUKA notation, so
\begin{equation}
  \begin{aligned}
R_1 \cup R_2 	& = & (+z_1 \: | +z_2)  \cup ( +z_3 \: | +z_4) \\
			& = & +z_1 \: | +z_2 |  +z_3 \: | +z_4
\label{eqn:setUnion}
\end{aligned}
\end{equation}
and
\begin{equation}
  \begin{aligned}
    R_1 \cap R_2 	& = & (+z_1 \: | +z_2) \cap ( +z_3 \: | +z_4) \\
			& = & +z_1 +z_3  \; | +z_1 +z_4 \; | +z_2 +z_3 \; | +z_2 +z_4\,.
  \end{aligned}
\label{eqn:setIntersection}
\end{equation}

Apart from the lattice feature, FLUKA has no sense of a volume hierarchy. Each body is placed with
translation, rotation and expansion geometry directives in global coordinates. A transformation from world
coordinates to a physical volume is built up by recursively applying daughter volume transformations
and  this is used to place FLUKA bodies. This  in practice is very similar to the procedure to create the
VTK visualisation already described in Section~\ref{sec:visualisation}.

In FLUKA every single point in space needs to be associated with one and only one region. This presents a problem
when converting Geant4 logical volumes to FLUKA regions, as the logical volume outer solid~$S_{\rm logical}$
needs to have the daughter  solids~$S_{{\rm daughter},i}$ subtracted. A solid which can be converted
to a region~$S_{\rm region}$ is then
\begin{equation}
S_{\rm region} =  S_{\rm logical} - S_{\rm daughter,1} - S_{\rm daughter,2} - S_{\rm daughter,3} \ldots
\label{eqn:logicalSubtraction}
\end{equation}
If a logical volume has a number of daughter volumes which are also possibly Boolean solids, then computing
$S_{\rm region}$ can become very complex because of Equations \ref{eqn:setDiff}, \ref{eqn:setUnion} and~\ref{eqn:setIntersection}.

Figure~\ref{fig:gdml-flair} shows an example conversion from GDML to FLUKA. The example is a
vacuum chamber with three ConFlat flange (CF) beam pipes connected to CF flanges. The top and bottom plates
have been removed to display the geometry more clearly. The model is formed of \verb|Box|,
\verb|Tubs| and Boolean operations of subtraction, intersection and union.

\subsection{Non-convex solid decomposition}
In general the BREP solids used in Geant4 include non-convex solids such
as polycones, extrusion solids, and twisted solids.  Converting these
solids to an equivalent FLUKA representation is problematic as non-convex
solids can only be created by the union of convex zones. Generating these
zones is non-trivial and these solids can be divided into two groups based
on the techniques required for the decomposition.  First are those where only a 2D polygonal
section needs to be decomposed, and these include polycones, polyhedrons
and extrusion solids.  The second group are those which require a full 3D
convex decomposition, and these include the twisted and tessellated solids.
This second group of non-convex solids are converted to CGAL Nef
polyhedra~\cite{cgal:hk-bonp3-20b} and decomposed to convex
polyhedra~\cite{cgal:h-emspe-20b}.

\subsection{Disjunctive normal form and degenerate surfaces}
FLUKA will typically decompose a region into disjunctive normal form (DNF)
at runtime, this normal form is characterised as the union of intersections and subtractions,
\begin{equation}
R = z_1 \; | \;z_2\;  | \; z_3 	\; | \; z_4 \dots
\end{equation}
Defining regions in terms of the DNF allows the rapid test of whether a point is inside that region. Testing each zone
$z_i$ of $R$ can terminate if a point is determined to be inside any of the zones $z_i$. In general
Equation~\ref{eqn:logicalSubtraction} does not have the form of a DNF. If there are many levels of logical-physical volume
placement, then recursive application of  Equation~\ref{eqn:logicalSubtraction} will create a nested set of
parentheses. There are some conditions where a general Boolean expression can yield an exponential
explosion of the final DNF. There are well known algorithms to convert logical expressions to its DNF.
\PYGEOMETRY{} can simplify parentheses from a region by creating a corresponding SymPy~\cite{10.7717/peerj-cs.103} Boolean
expression and using the \verb|to_dnf| method.  Further simplification of the CSG tree leverages \PYGEOMETRY{}'s
meshing capabilities combined with CSG pruning algorithms based
on~\cite{pruning}.  FLUKA will by default try to expand all regions to
their DNF at runtime, which inevitably can result in the sort of exponential explosion already mentioned.  Until version 4.0, if FLUKA identified
such an explosion, it would terminate the expansion, report an error and exit, thus making such models impossible to run.  As of version 4.0, however, this
expansion can be disabled, and the tracking algorithms will walk the CSG trees verbatim, albeit at some tracking efficiency cost.  Therefore, most
output generated from the GDML to FLUKA conversion described here can only be used with FLUKA 4.0.

\subsection{Materials}
\PYGEOMETRY{} converts GDML/Geant4 materials to FLUKA \verb|MATERIAL| and \verb|COMPOUND| cards.
Geant4 has a class \cpinline{G4Material} to assign material state (density, physical state, temperature and pressure)
to a logical volume. \cpinline{G4Material} has two main constructors, the first where an atomic number is supplied and
the second is when \cpinline{G4Element} instances and relative atomic or mass abundances are provided. The {\em simple}
\cpinline{G4Material} is converted to a \fluka{material} card, whilst the {\em element} \cpinline{G4Material}  is converted to a
\fluka{compound} card. There are similar issues when converting \cpinline{G4Element} to FLUKA, as \cpinline{G4Element}
can either be simple, i.e. defined only by atomic number and mass, or composite and defined by an admixture of
relative abundances of isotopes. A similar mapping is performed so that if a \cpinline{G4Element} is simple it is
directly converted to a FLUKA \fluka{MATERIAL} card, and the {\it isotope} \cpinline{G4Element} is converted to a \fluka{compound}
card. Geant4 also defines a set of standard materials~\cite{Geant4MaterialDB} or compounds from the US National Institute
of Standards and Technology (NIST), so a user can, for example, simply
specify the name \pyinline{G4_STAINLESS-STEEL}. \PYGEOMETRY{}  contains a matching database and creates the appropriate FLUKA
cards from these names during the conversion. This database is updated when
necessary by running a small Geant4 program to output the appropriate material data.

\subsection{Discussion}
Overall the conversion to FLUKA input format from GDML is quite advanced
and stable. Relatively large experimental simulations have been converted
from GDML to FLUKA and have been used to produce simulation results. The
conversion described still requires a user to understand how geometry is
specified in both Geant4 and FLUKA. For example, as has been stated, a
Geant4 mother volume will, when translated to a FLUKA region, have all of
its daughters subtracted from its respective FLUKA region. If those daughter volumes, when
translated to FLUKA, consist of combinations of many FLUKA bodies, the
mother region can become very complex.  This complex
region will generally not be in disjunctive normal form and is inefficient
when viewed in FLUKA's graphical user interface (GUI), flair, and used for
simulation in FLUKA.  Users who wish to minimise this should ensure that
there are not large numbers of daughter volumes at the same level in the
hierarchy, unless corresponding to solids which translate to simple
primitives such as boxes.  Complex daughter volumes consisting of solids
that are complex Booleans should be placed inside solids that map
to primitive FLUKA bodies.  This ensures that expanding to the DNF is
computationally tractable, and in doing do FLUKA's tracking will be faster.
This is not a drastic constraint on the conversion functionality, however,
as this is actually typically how one would implement geometries in FLUKA
regardless.

With the above in mind it is perhaps preferable for a user creating
geometry from scratch to use \PYGEOMETRY{} and these conversion tools
as part of the process.  This would allow a user to target multiple codes with
a single source description with minimal additional effort.  However, there
are still a  outstanding technical issues with the conversion, which are
discussed in the rest of this section.

Replica, division and parametrised placements are not currently implemented
and will be added in a future release. In Geant4 it is possible to create scaled
solids or placements with reflections (referred to as \emph{scale} in Geant4). FLUKA
rototranslations do not support reflections and implementing reflections
requires transformation of the body definitions. This is not yet implemented
in the current version of \PYGEOMETRY{}.

In general, recursive application of Equations \ref{eqn:setDiff},
\ref{eqn:setUnion}, \ref{eqn:setIntersection}
and~\ref{eqn:logicalSubtraction} can result in very complex regions when
converting from Geant4/GDML to FLUKA. The complexity of the final region
expression can be compounded if transformed to DNF. The final region
Boolean expression in DNF can be simplified by using the meshes of the
intermediate parts, only retaining terms which do not contribute to the
overall definition of the geometry (e.g. subtracting with a solid that
does not overlap any other solid), using an algorithm similar to the one
outlined in~\cite{pruning}.

It is possible given the GDML to FLUKA conversion algorithm described in
this paper that coplanar overlaps exist in the FLUKA geometry.  In Geant4
there is no connection between surfaces used to specify one logical volume
solid and another logical volume solid. For example if a logical volume
solid shares a face with one its daughter volume solids, and those faces
when translated to FLUKA are expressed in terms of half-spaces for example,
those half-spaces would be duplicated in the final FLUKA file.  It is
possible to remove obvious degeneracies but this is complicated by
placements of bodies. Every FLUKA body with an arbitrary number of
associated transformations (rototranslations, rotations or expansions)
provided by geometry directives can be re-expressed purely in terms of a
body (possibly of a different type) without any transform. A simple example
of this is the XYP and PLA, it is simple to transform an XYP into a PLA
with an appropriate rotation. This {\em normal form} can be used to test
for approximate equality between bodies that are equivalent whilst
accounting for any transforms.  Approximate equality is required as
multiple applications of rototranslations will accrue numerical rounding
errors such that true equality is not maintained.  This would allow for the
removal of degenerate surfaces removing potential coplanar overlaps and
also reduce the final converted file size.

The twisted primitive solids need to be decomposed
into a union of convex solids. This decomposition does not always
succeed or produces a far from optimal number of convex solids.
An alternative to implementing these solids is to approximate each
layer of the twisted solid as a union of tetrahedra. A similar problem
exists for tessellated solids, which in general
are non-convex and need decomposition into convex hulls. An alternative
to the computationally expensive convex decomposition is to
create a region formed from the unions of tetrahedra. There are numerous
algorithms for tetrahedralisation of surface meshes in both CGAL and TetGen~\cite{tetgen}
and these will be implemented in a future release of \PYGEOMETRY{}. Even if a stable and general
method for converting tessellated solids exists, it is not efficient to define
tessellated objects in this way.  Memory, body or zone limits might
be reached in FLUKA, thus possibly limiting the size of CAD or STL models which
might be loaded.

\section{STL and CAD to GDML conversion}
STL and CAD conversion are closely related. In both cases the solid
(in the case of STL) and solids (in the case of CAD) are converted to
tessellated solid(s). STL is a relatively simple file format that can
be loaded using pure Python. As STL files typically only contain a single
solid the \pyinline{stl.Reader} provides a single solid and not a logical volume
as with the other file readers.
\begin{figure}
\begin{center}
\includegraphics[width=0.9\columnwidth]{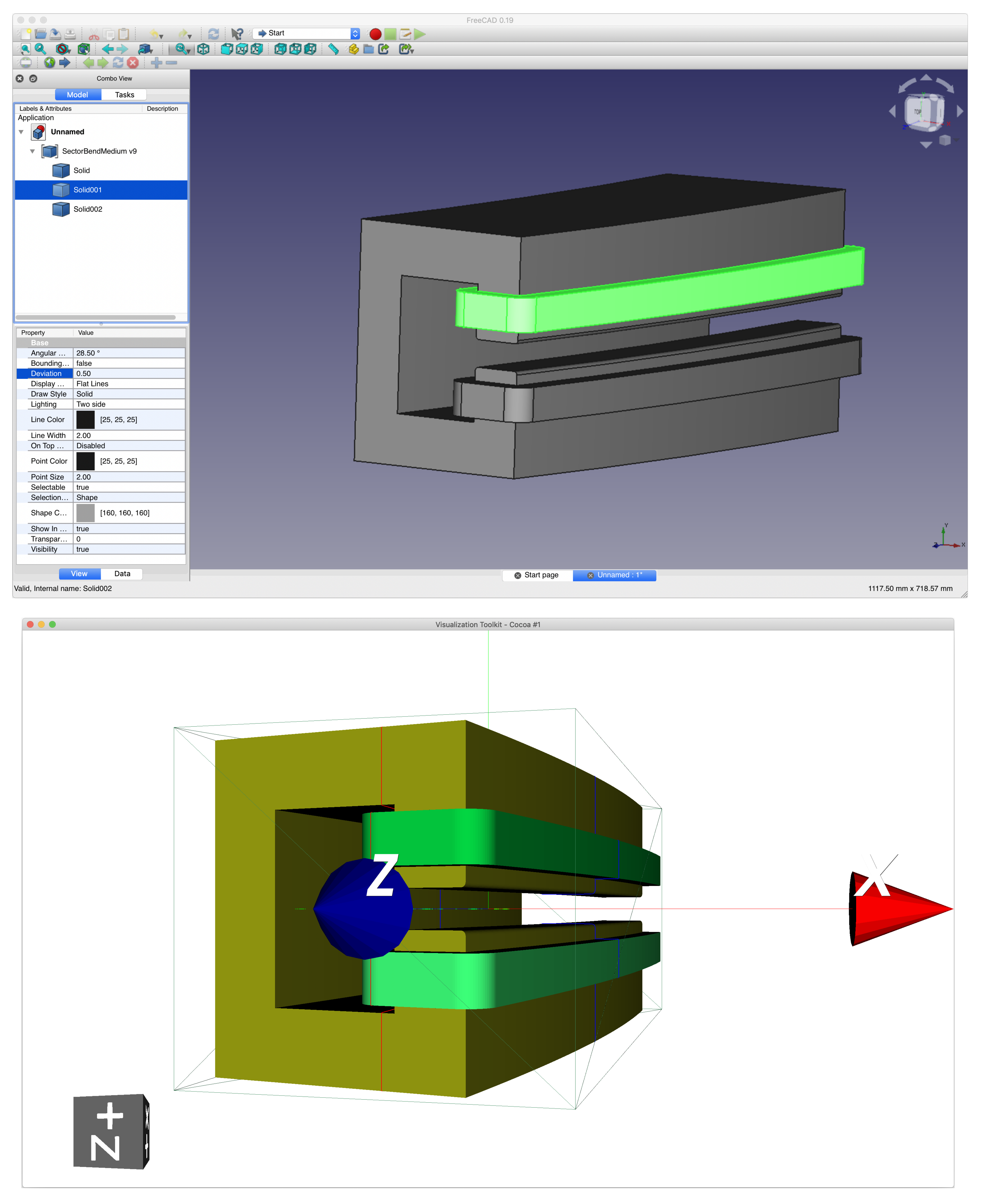}
\caption{Example conversion of a simple CAD (STEP) geometry (sector bend
  dipole electromagnet) to GDML, above: the original model in FreeCAD,
  below: the GDML geometry viewed using \PYGEOMETRY{}. }
\label{fig:cad-gdml}
\end{center}
\end{figure}

STEP and IGES files can be loaded into \PYGEOMETRY{}, via an interface
based on FreeCAD~\cite{FreeCAD}. FreeCAD is an open source CAD/CAM program,
which in turn is based on OpenCASCADE. FreeCAD allows for scripting in
Python and acts as a simple-to-use interface to OpenCASCADE.  A STEP CAD
file could be considered as a hierarchical tree of parts and \emph{part
  assemblies}, where a part assembly is a collection of \emph{part
  features}. A part feature can be used to create a triangular mesh which
can be used to instantiate a \PYGEOMETRY{} \pyinline{TessellatedSolid}. The
placement of the part feature is extracted from the STEP file and used to
create an appropriate physical volume. Assignment of materials and
visualisation attributes must be performed by the user after conversion to
GDML as it is rarely the case that CAD/CAM packages include the detailed
information required for MCRT codes. The loading of STEP files in
\PYGEOMETRY{} is demonstrated in Listing~\ref{lst:pythonCADLoading}.

\begin{lstlisting}[caption={A simple \PYGEOMETRY{} Python script to load a STEP file.},label={lst:pythonCADLoading}, language=Python]
import pyg4ometry.freecad as freecad

reader = freecad.Reader("CadFileName.step")
reader.relabelModel()
reader.convertFlat()
logical = reader.getRegistry().getWorldVolume()
\end{lstlisting}
Compared to other file readers, two additional steps are required: \pyinline{relabelModel} and \pyinline{convertFlat}. CAD model
part names can contain characters which are not allowed in Python
dictionaries so need to be replaced by using \pyinline{relabelModel}.
CAD models might also have a hierarchy of parts and assemblies, these are
converted without this structure by using \pyinline{convertFlat}.
In general there is no requirement to avoid geometric overlap  of parts in
a CAD file. This will result in overlaps between the converted tessellated
solids. This is avoided by shrinking each solid by computing the normal ${\bf n}$  for each
vertex ${\bf v}$ and shifting its position by $\epsilon {\bf n}$, so that
the new vertices are positioned at ${\bf v} - \epsilon {\bf n}$. The degree of shrinkage
is user-controllable.

An example geometry representing a dipole electromagnet, consisting of three parts was created in Autodesk Fusion 360
and saved as a STEP file, the \PYGEOMETRY{}-produced output is shown in Figure~\ref{fig:cad-gdml}.

\section{Complete simulation example}
\begin{figure*}
\begin{center}
\includegraphics[width=1.0\textwidth]{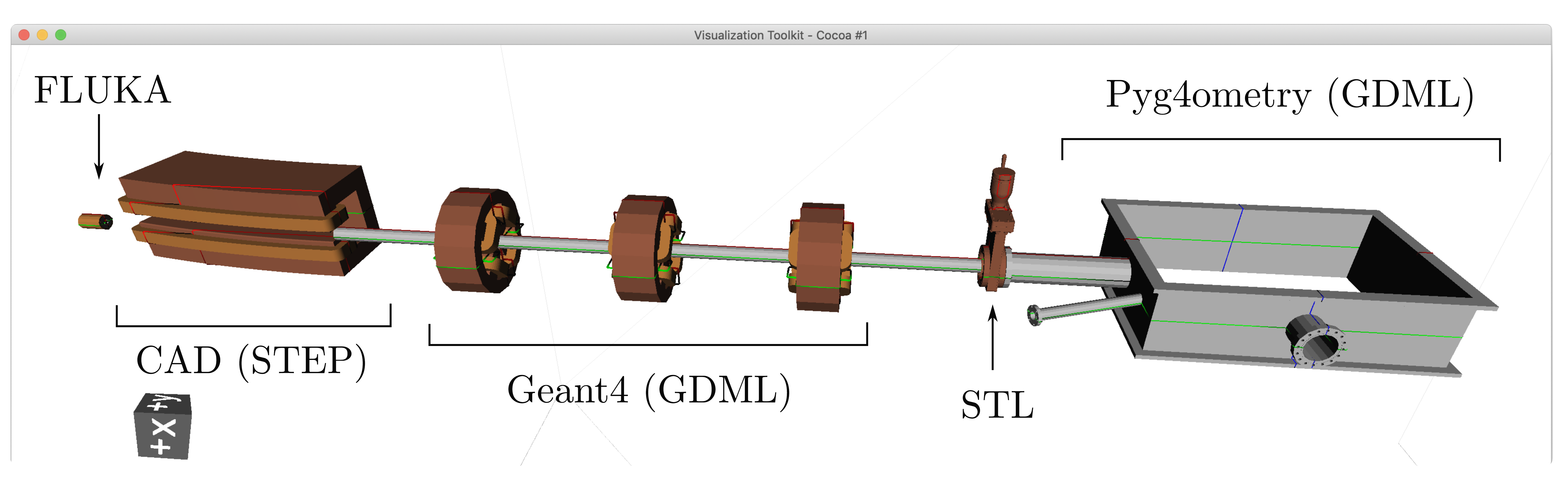}
\caption{Complete composite example using \PYGEOMETRY{} showing a model with geometry sourced from a range of different formats.}
\label{fig:model}
\end{center}
\end{figure*}

\PYGEOMETRY{} is designed to be as flexible as possible and offer the user a wide range of usage styles, input files
and workflows. A fictitious beamline was created to demonstrate the capabilities of \PYGEOMETRY, this creates a composite
\emph{scene} which consists of geometry sources from the different formats described in this paper. The beamline consists of a vacuum chamber
(modelled in \PYGEOMETRY{}), a vacuum gate valve (STL from the manufacturer), a triplet of quadrupole magnets (exported to GDML
from BDSIM), a sector bend dipole electromagnet (created in Autodesk Fusion 360) and finally a Faraday cup (FLUKA geometry designed in
flair). Each different file is loaded using the appropriate \PYGEOMETRY{} \pyinline{Reader} class and then placed as a physical volume. The final composite
geometry is shown in Figure~\ref{fig:model}. It must be noted that when this complete geometry is written to GDML and loaded into
Geant4 it cannot be visualised with anything but the ray tracer due to limitations in the OpenGL visualisation in Geant4. Another
important note is that this geometry cannot be converted to FLUKA as it contains tessellated solids (both the STL gate valve and
dipole magnet).

Having geometry wrapped in a suitable API allows a wide range of processes to be performed simply and programmatically. The benefits of the API are
particularly apparent when needing to process large amounts of geometry efficiently and precisely. Possible transformations include merging registries,
removing volumes (defeaturing), editing solid parameters, changing logical volume materials and converting logical volumes to assembly volumes.

The merging of registries and removing of volumes is required to create the example shown in Figure~\ref{fig:model}.
Each sub-component is stored in a separate registry and these have to be combined without any GDML tags clashing.
The FLUKA to GDML conversion creates logical volumes which are not generally required in Geant4, so for example
the air surrounding the Faraday cup, which is needed to specify a FLUKA geometry is converted to GDML but can be
safely removed. Examples of other workflows or geometry manipulation processes can be found in
the \PYGEOMETRY{} online manual.

\section{Quality assurance}
The source and manual code for \PYGEOMETRY{} is stored in a Git repository (\url{https://bitbucket.org/jairhul/pyg4ometry}),
where a public issue tracker for users is hosted to report problems or bugs
with the code. The Python source is documented throughout using docstrings, which
are also used to generate an API reference using Sphinx, a
Python documentation generator
(\url{http://www.pp.rhul.ac.uk/bdsim/pyg4ometry/}).  This
automatically-generated documentation is supplemented with
developer-written documentation in the form of examples and tutorials,
again using Sphinx. \PYGEOMETRY{} uses mature packages
available for Python as dependencies. \PYGEOMETRY{} has  two sub-packages
that require compilation: \pyinline{pyg4ometry.pycsg} in Python and
\pyinline{pyg4ometry.pycgal} in C++.
\PYGEOMETRY{} package dependencies and extensions are easily installed using
\pyinline{setuptools}. All aspects of \PYGEOMETRY{} are routinely checked
using software tests, of which there were 543 at the time of writing,
resulting in 84\% code coverage. The tests also serve as minimal examples to help users understand the code operation.

\section{Conclusions and discussion}
The authors believe that tools to quickly create geometry, either from scratch or by conversion,
for Monte Carlo particle transport programs
will save significant amounts of time and user effort and will ultimately
yield more accurate simulations. \PYGEOMETRY{} is a relatively complete implementation of a geometry
creation tool, and whilst heavily internally based on Geant4 and GDML, it can have utility for users of all MCRT
codes. \PYGEOMETRY{} can clearly be extended to other formats or applications. Presently it provides a coherent and
uniform interface to existing tools and utilities, and by using the Python programming language
 the programmatic control of geometry creation or modification is possible. This approach allows the integration of
other available tools~\cite{DavisNIMA915-65} into a unified workflow.

Users should be aware of issues with \PYGEOMETRY{}. The conversions between CAD/STL, GDML, and FLUKA
cannot be considered bidirectional. For example, Geant4 tessellated solids cannot be converted easily to FLUKA which does not
have a convenient way of representing this geometry. In general a user would be unwise to attempt to convert a very large
geometry from one format to another, but should instead concentrate on smaller conversions of constituent parts, 
and then using those parts to supplement larger models. Workflows should focus
on the generation of a primary format and then create conversions to other formats as the need arises. This paper outlines
the creation of geometry using \PYGEOMETRY{}, and whilst that geometry is subsequently loaded into Geant4, flair and FLUKA, detailed
studies of the MCRT simulation performance is beyond the scope of this publication and will be addressed in the future.

There are many output format extensions that can be considered for \PYGEOMETRY{}.
Geant4 geometry is primarily created by writing C++ programs, so an output writer that
converts the \PYGEOMETRY{} in-memory representation to C++ will allow rapid geometry
modelling and inclusion of the geometry into an existing Geant4 application. This is not
implemented in the current version but could be relatively quickly implemented for users that
require this functionality. At present \PYGEOMETRY{} supports reading and writing
Geant4 (GDML) and FLUKA files but could be extended without significant effort to other MCRT codes
like MCNP.

There are more complex extensions that can be considered for inclusion into \PYGEOMETRY{}.
The meshes created by \PYGEOMETRY{} are generally of very high quality and can be used for a
wide range of applications. An idea already being developed is the export of the geometry mesh data to
data formats used in augmented or virtual reality software to create interactive visualisations of MCRT
simulations.  Triangular meshes also have applications for GPU-accelerated photon tracking in
optical photon based particle physics detectors. ParaView/VTK are becoming standard software for 
complex 3D visualisation and the ability to write geometry to formats readily loaded and manipulated 
by these programs will significantly aid the presentation of geometry along with the results of MCRT 
simulations.

\PYGEOMETRY{} is principally a toolkit but various visualisation and user interface extensions would
significantly aid geometry creation workflows. A graphical user interface would enable a user without any
 programming experience to create geometry for MCRT simulations and expand the number of potential users.
\PYGEOMETRY{} has been designed to interface with a GUI in a relatively straightforward manner. The VTK visualiser currently
limits the display of very large models as geometry instances are replicated as opposed to reused in visualisation.  However,
this could be drastically improved in a future release.

The conversion which would most dramatically enhance the creation of MCRT geometry is CAD to Geant4 or FLUKA
without use of a triangular or tetrahedral mesh. There are existing approaches to decompose BREP solids to Geant4 and
FLUKA-like CSG geometry~\cite{WangNuclSciTech31-82-2020, LuFusionEngineeringAndDesign124-2017}.
The FreeCAD/OpenCASCADE interface combined with the Geant4 and FLUKA Python API in \PYGEOMETRY{}
will allow for the creation of CAD BREP decomposition algorithms. There are Python-based CAD modelling tools like
CadQuery~\cite{cadquery} which allow the creation of models using pure Python which should allow the conversion of GDML to STEP.

There is a strong relationship between \PYGEOMETRY{} and Geant4 and to a lesser extent between
\PYGEOMETRY{} and FLUKA. \PYGEOMETRY{} can be used as a testing ground for ideas prior to
implementation in Geant4 or FLUKA. An example of this is the VTK visualisation system implemented in
\PYGEOMETRY{}, which could be used in Geant4 to render Boolean solids which frequently
fail in the Geant4 OpenGL viewer, despite being otherwise perfectly valid constructs.
This would involve using CGAL meshing in the \pyinline{G4Polyhedron} class. %There is also

\PYGEOMETRY{} is already proving to be a useful tool for geometry conversion, creation and manipulation.
There are numerous international researchers and research groups already using the code for their particular applications.
The users are focused in accelerator physics, but \PYGEOMETRY{} could find application in any scientific
area where MCRT simulations are needed, for example particle physics, space environment and medical physics.
The authors welcome contributions, extensions and bug fixes as well as suggestions for larger collaborations.

\section{Acknowledgements}

The development of \PYGEOMETRY{} has received funding from Science and
Technology Research council grant ``The John Adams Institute for
Accelerator Science'' ST/P00203X/1 through the John Adams Institute at
Royal Holloway, and Royal Holloway Impact Acceleration Account. Gian Luigi D'Alessandro (CERN/JAI)
provided geometry and tested multiple converted beamlines. Multiple
undergraduate and masters students have contributed to the code for their
degree project work: Simon Williams, Benjamin Shellswell and Joshua Albrecht.
STB has benefited from discussions with, and the insights of, Vasilis Vlachoudis
(CERN) regarding FLUKA geometry and flair.  Cédric Hernalsteens (ULB/CERN) and
Alistair Butcher (RHUL) also made useful contributions to the code. Don Boogert
provided valuable insights on meshing and modern computer graphics programming.



%% References


\bibliographystyle{elsarticle-num}
\bibliography{pyg4ometry}





% \end{thebibliography}

\end{document}